# APPLICATION OF EXECUTIVE INFORMATION SYSTEM FOR COVID-19 REPORTING SYSTEM AND MANAGEMENT: AN EXAMPLE FROM DKI JAKARTA, INDONESIA

Verry Adrian, Jakarta Department of Health, dr.verry@gmail.com

Intan Rachmita Sari, Jakarta Department of Health, intanrachmita@gmail.com

Hardya Gustada Hikmahrachim, Jakarta Department of Health, hardyagustada@yahoo.com

**Abstract:** SARS CoV-2 infection and transmission are problematic in developing countries such as Indonesia. Due to the lack of an information system, Provinces must be able to innovate in developing information systems related to surveillance of SARS CoV-2 infection. Jakarta Department of Health built a data management system called Executive Information System (EIS) of COVID-19 Reporting. EIS aimed to provide actual data so that current epidemiological analysis is accurate. The main idea of EIS is to provide valid and actual information to stakeholders, which can then be presented in the form of a dashboard. EIS is utilized to push data flow and management for rapid surveillance purposes. This could be the first time in Indonesia that a system reports near-actual data of nearly half a million people daily using an integrated system through a transparent system. The main data presented is important to monitor and evaluate COVID-19 transmission is the cumulative case dan daily case number. Data in EIS also can offer data geographically so that a more detailed analysis could be done. EIS's data and the dashboard help the government in pandemic control by presenting actual data on bed occupancy and availability across hospitals, especially isolation wards. Stakeholders, academic institutions should utilize EIS data and other elements to help Indonesia fight COVID-19.

## 1. INTRODUCTION

The first coronavirus disease (COVID) 19 case in Indonesia was reported in March 2020, specifically in Depok, a city near Jakarta. Since that, massive infection and local transmission are unstoppable in Indonesia. In the middle of March 2021, around 1.5 million people were infected by SARS CoV-2 virus in Indonesia, with currently 122.000 active cases and a 2.7% mortality rate. The highest cases were found in Jakarta. Yet, Indonesia had not overcome COVID-19.

Due to the lack of an information system, provinces are pushed to do any innovation related to SARS CoV-2 infection surveillance. Daily data reported are mandatory for every province, consist of new cases, recovered cases, and death cases. Those data were then collected to be summarized as a daily national report.

To provide valid and reliable data, Jakarta Department of Health had been developing a data management system called Executive Information System (EIS) of COVID-19 Reporting. EIS aimed to provide actual data so that current epidemiological analysis is accurate. Any decision for pandemic control should be based on high-quality data. This paper aimed to present the design and outcomes of EIS utilization in Jakarta. This system had been build at the beginning of 2018. The main idea of EIS is to provide valid and actual information to stakeholders in the form of a dashboard.

This idea is growing as demand increases during the pandemic, especially when COVID-19 case reporting emerged.

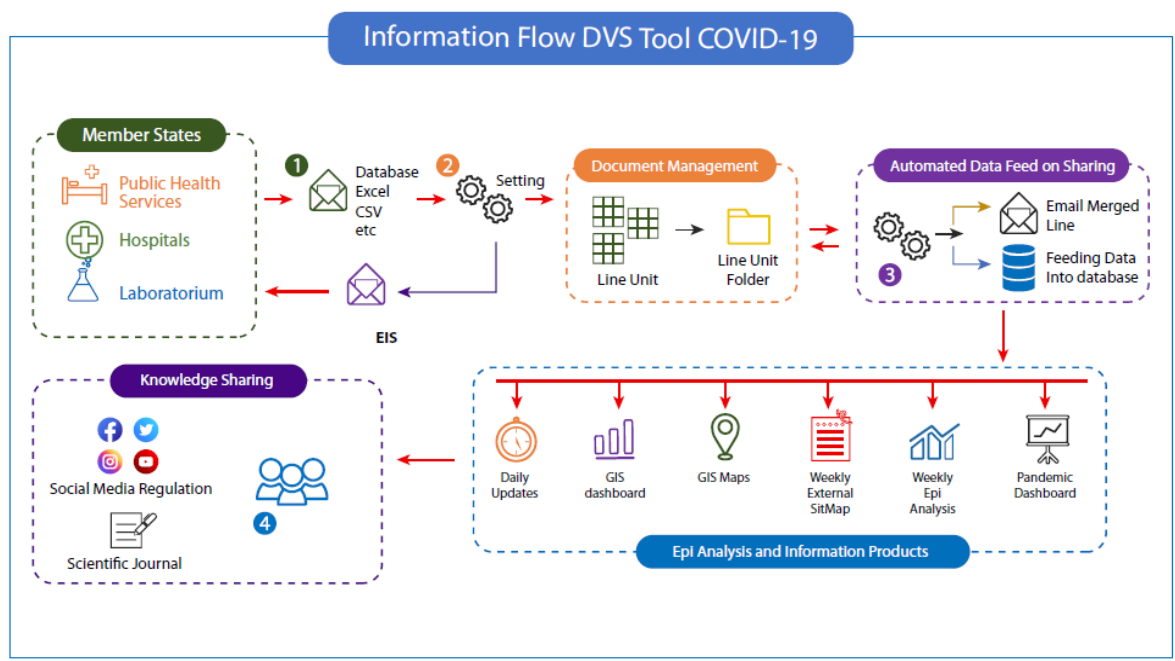

**Figure 1. Information flow in EIF build by Jakarta**

During the COVID-19 pandemic in Jakarta, EIS is utilized to push data flow and management for rapid surveillance purposes. As seen in Figure 1, Jakarta's data sources are public health services, hospitals, and laboratories. Both patient data and the result of PCR data are reported by those healthcare facilities by using an online database, excel, csv, and other tools. Data were then processed into EIS so that further steps can be integrated. Document management primarily uses line unit and line unit folder and is brought to automated data feed on sharing. Epidemiologic analysis and presentation (information products) of EIS in Jakarta are daily COVID updates (will be explained later), GIS dashboard and maps, weekly external SitMap and Epi Analysis, and also pandemic dashboard.

This data would then be utilized by epidemiologists and public health analysts for pandemic-related policy, writing of scientific reports, and knowledge sharing via social media. Using this system, actual data were presented and accessible for all elements in Jakarta. Due to a large amount of data incoming across the province, the EIS system optimizes data verification by collaboration with the civil registration department to prevent data duplication, mostly using a single civil registration number (Nomor Induk Kependudukan or NIK) (Fig 2). Although it still needs improvement, this scheme prevents potential data duplication during COVID-19 case reporting. The significant difference of data collection and analysis before and after the establishment of EIS is presented in Table 1.

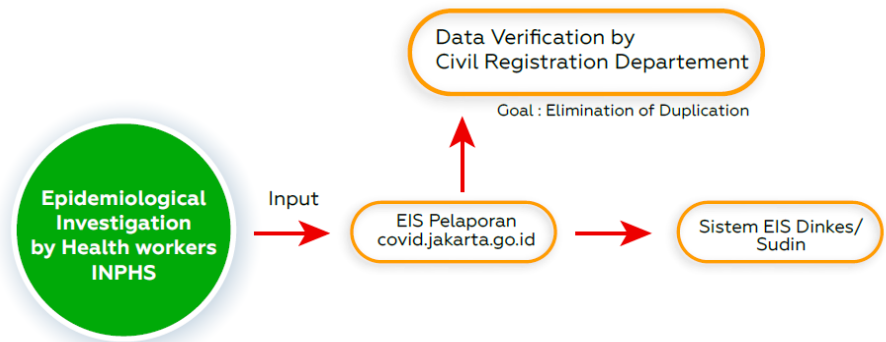

Figure 2. Concept of Data Management during COVID Pandemic

Table 1. Difference in data collection to data analysis before and after the establishment of EIS

| Aspects | Before | After |
| --- | --- | --- |
| Data collection | Written in paper-based form, then recapitulated manually. | Direct input into system and automatic data summary by system. |
| Reporting | Manually recapitulated data sent into stakeholders in higher level for data input into system. | Data connected via API, directly visualized in dashboard. |
| Time consumption | Needs longer time due to manual processing from data input to data analysis. | Less time needed for data analysis due to interconnected system. |
| Data quality | Moderate quality data due to incomplete data, duplication, and human error during data collection process. | Better data quality due to mandatory variable input in EIS system. Not possible for data duplication due to similar data will be automatically merged. Accumulative calculation is done by system to prevent human error during data calculation. |
| Data analysis | More difficult step for data analysis and data visualization due to manual process. | Automatic data analysis and visualization in the form of table, graph, or any presentation that been set before. |
| Data sharing | More difficult due to manual sending data to every stakeholder. | Data is accessible both for stakeholders and public as presented in the dashboard. |

## 2. DATA PRESENTATION AND VISUALIZATION

This EIS COVID-19 Reporting system's main objective is to provide data for Jakarta Provincial Government and then assisted by the Department of Communication, Informatics, and Statistics; data will be reprocessed and visualized in the form of a dashboard for use by stakeholders, academic institutions, and public consumption. Data should be regularly updated and give a lot of useful information. This is probably the first time in Indonesia that a system could report near-actual data of nearly half a million people daily using an integrated system through a transparent system. EIS is the back end of this system information, while the front end is presented on Jakarta COVID-19 website. Current dashboard that presented at corona.jakarta.go.id:

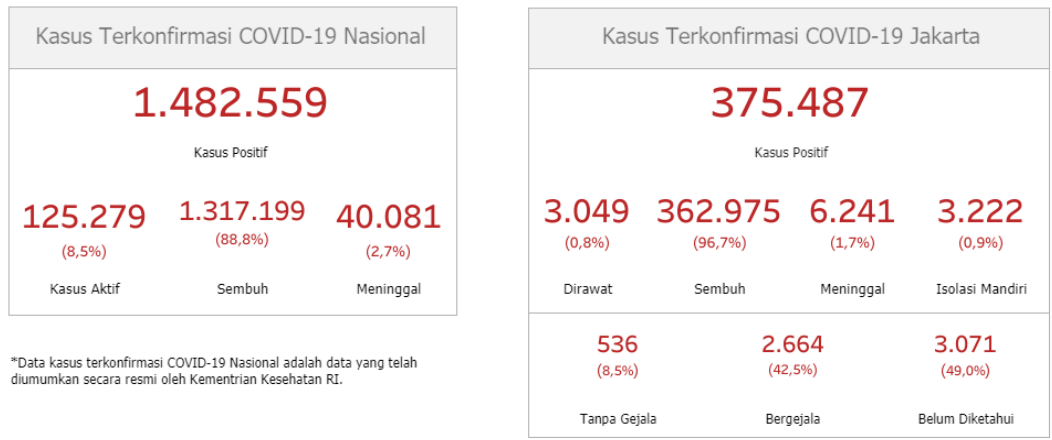

Figure 2. Daily Summary of Cummulative COVID-19 Cases in Indonesia (left) and Jakarta (right). Kasus terkonfirmasi = Confirmed cases; kasus positif = positive cases; kasus aktif = active cases; sembuh = recovered; meninggal = death; dirawat = hospitalized; isolasi mandiri = self-isolation; tanpa gejala = asymptomatic; bergejala = symptomatic; belum diketahui = unknown.

The main data presented that is important to monitor and evaluate COVID-19 transmission is cumulative case dan daily case number. Using EIS, Jakarta could present a more specific data compared to national data by reporting self-isolation case, asymptomatic cases, and symptomatic cases. Those proportion is important to estimate how severe this disease are among people of Jakarta.

Figure 3. Detailed information about suspect (suspek), probable, travelers (pelaku perjalanan), close contact (kontak erat), and discarded cases (cumulative). Selesai isolasi = finished self-isolation; isolasi di rumah = self-isolation at home; isolasi di RS = isolation at hospital; meninggal = death.

Data above present a stratification analysis about COVID-19 in Jakarta, especially about suspected cases, probable cases, travellers cases, close-contact cases, and discarded cases. This information is crucial as Indonesia yet do not have an optimal testing number and strategies and near half of them are conducted in Jakarta. Cumulative number of self-isolation compared to hospital admission can be an information to be used for policy making about Jakarta capacity to fight for this pandemic.

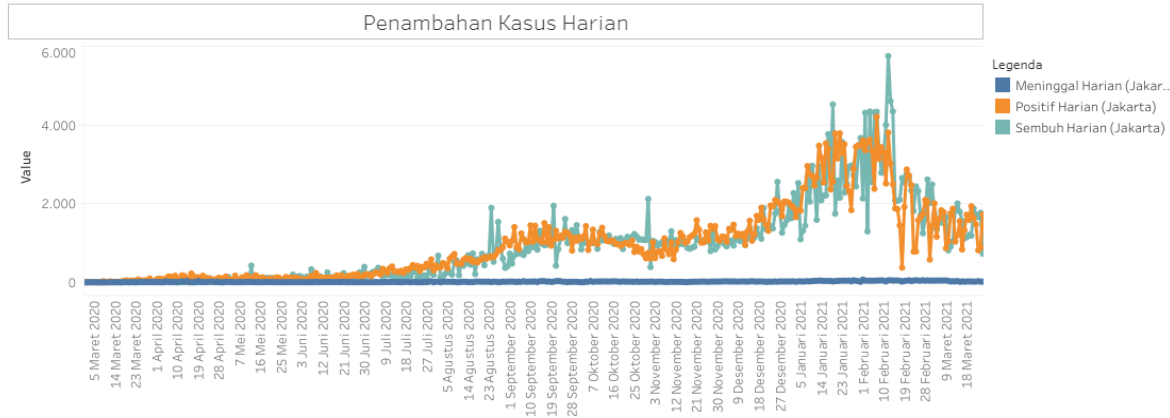

Figure 4. Daily New (Penambahan Kasus Harian) COVID-19 Case and Mortality Report. Meninggal harian = death case daily; positif harian = new positive case daily; sembuh harian = recovered case daily

Another important role of EIS is that it can visualize an actual report of daily new cases and its trend (Fig 5). This graphic has been adapted based on other international health organization report such as US CDC and UK NHS. Accumulative cases could also be accesses freely as it also presented at COVID-19 website after being processed in EID (Fig 6).

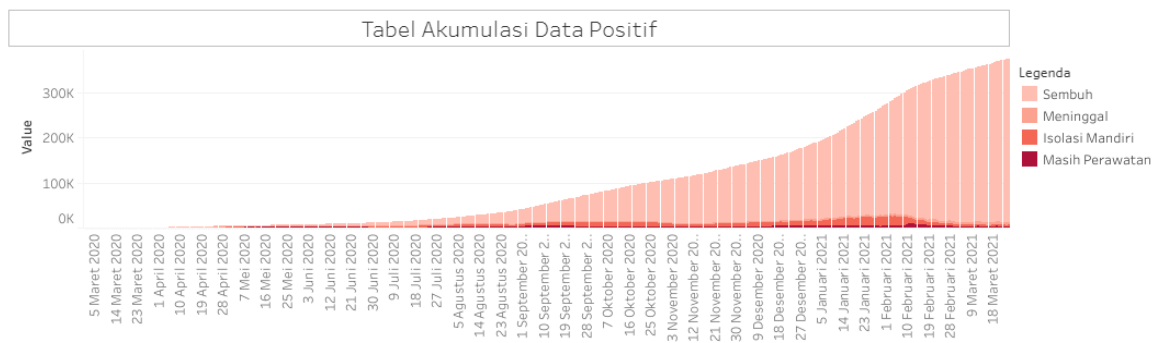

Figure 5. Cumulative (Akumulasi) Data on COVID-19 Active (Masih perawatan), Death (meninggal), Self-Isolation (isolasi mandiri), and Recoved Cases (Sembuh)

As seen in Fig 7, EIS also can report the daily positivity rate of COVID-19 diagnostic test in Jakarta. The trend of specimen tested and positivity rate helps epidemiologist in deciding for the further strategies. Any strategies related to pandemic control would not be explained in detail in this paper.

| Tanggal | Jumlah Orang di Test | Orang Positif Harian | Orang Negatif Harian | Positivity Rate Kasus Baru Harian | Total Spesimen di Test | Positif | Negatif | Positivity Rate Spesimen Harian |
|---|---|---|---|---|---|---|---|---|
| 12/03/2021 | 10.625 | 1.034 | 9.591 | 9,7% | 14.220 | 3.237 | 10.983 | 22,8% |
| 11/03/2021 | 12.062 | 1.873 | 10.189 | 15,5% | 13.865 | 2.586 | 11.279 | 18,7% |
| 10/03/2021 | 13.016 | 1.754 | 11.262 | 13,5% | 18.264 | 4.302 | 13.962 | 23,6% |
| 09/03/2021 | 11.520 | 1.040 | 10.480 | 9,0% | 17.709 | 4.373 | 13.336 | 24,7% |
| 08/03/2021 | 10.823 | 867 | 9.956 | 8,0% | 15.215 | 3.357 | 11.858 | 22,1% |
| 07/03/2021 | 11.041 | 1.783 | 9.258 | 16,1% | 5.460 | 1.356 | 4.104 | 24,8% |
| 06/03/2021 | 7.524 | 1.834 | 5.690 | 24,4% | 10.327 | 3.116 | 7.211 | 30,2% |
| 05/03/2021 | 8.581 | 1.616 | 6.965 | 18,8% | 11.875 | 3.639 | 8.236 | 30,6% |
| 04/03/2021 | 10.041 | 1.159 | 8.882 | 11,5% | 15.761 | 3.477 | 12.284 | 22,1% |
| 03/03/2021 | 13.655 | 2.008 | 11.647 | 14,7% | 17.590 | 3.883 | 13.707 | 22,1% |
| 02/03/2021 | 19.899 | 1.437 | 18.462 | 7,2% | 25.673 | 6.521 | 19.152 | 25,4% |

Figure 6. Positivity Rate and Other Related Data on COVID-19 Testing in Jakarta. Tanggal = date; jumlah orang di test = people tested; orang positif harian = daily new positive cases; orang negative harian = daily negative cases; kasus baru harian = new active cases daily; total specimen di test = number of specimen tested; positivity rate specimen harian = daily positivity rate.

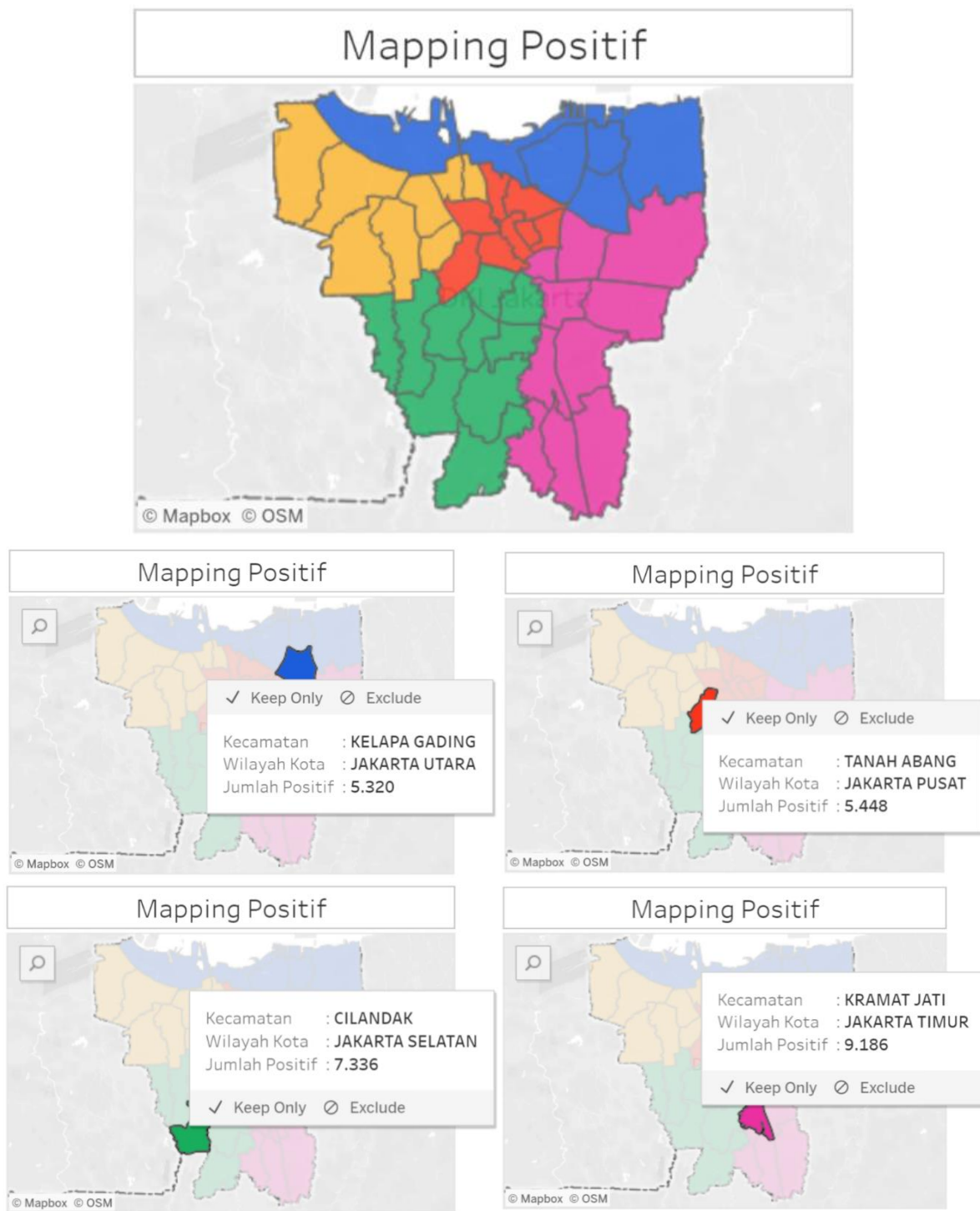

Figure 7. Positive COVID-19 Cases based on Administrative Regions. Positif = positive; Kecamatan = district; Wilayah Kota = administrative city; jumlah positif = positive cases.

Data in EIS also can present data geographically so that a more detailed analysis could be done. Data reported as cumulative (Fig 8) or daily new cases (Fig 9). New cases reported in detailed information such as age, gender, and hospitalization status.

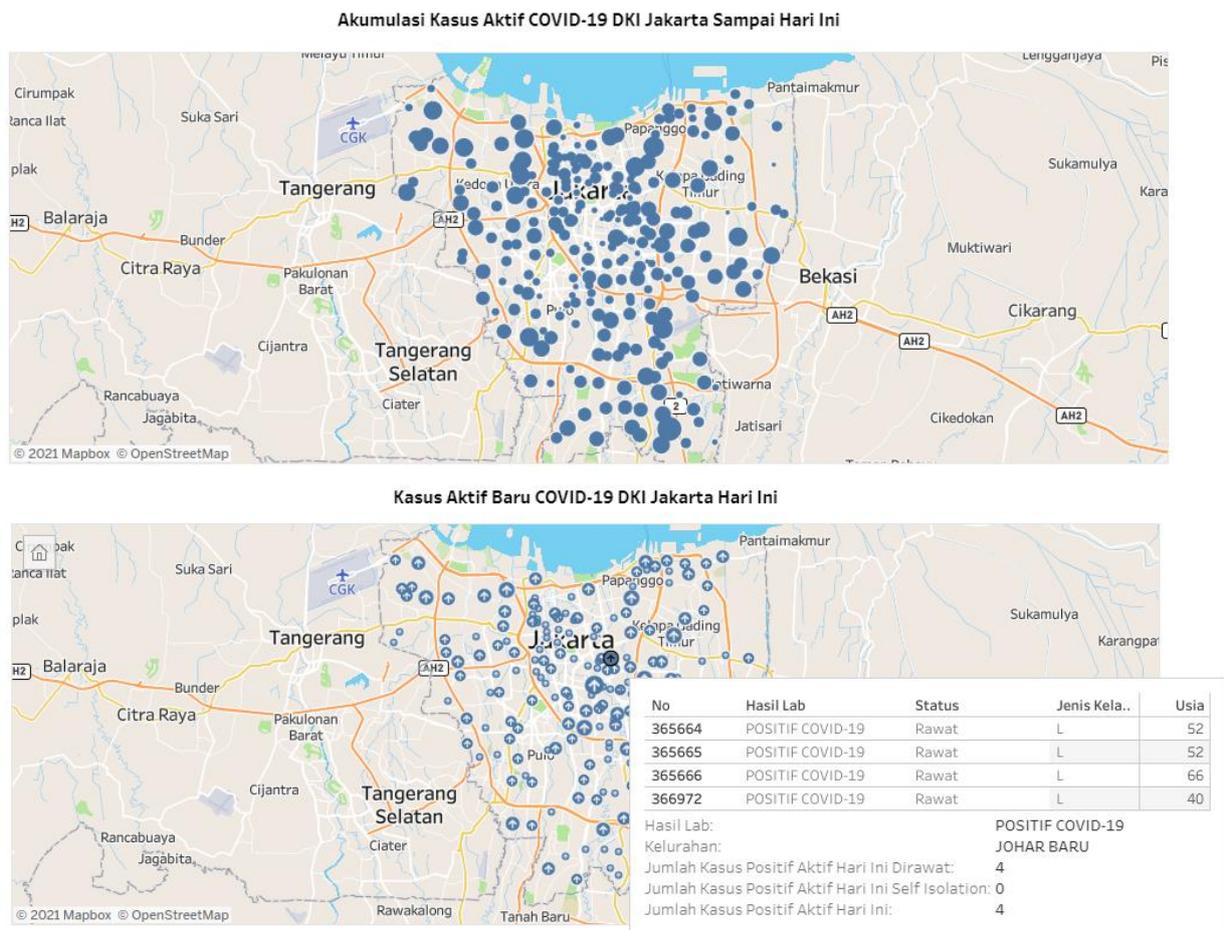

Figure 8. Mapping of New Positive Case Weight by Case Number (upper) and
by Basic Demographic Data (lower)

Information at EIS dashboard that can be accessed by public are:

```
 1. National COVID confirm cases – cumulative (active, recovered, and death
 cases)
 2. Jakarta COVID confirm cases – cumulative:
      i. Hospitalized
     ii. Recovered
    iii. Death
     iv. Self-isolation
      v. Asymptomatic
     vi. Symptomatic
    vii. Unknown symptom status
 3. Mapping of suspected, close-contact, and positive cases in Jakarta
 4. Crosstabulation data on gender and age group
 5. Daily new suspected, close-contact, and positive cases
 6. Comparison between national and Jakarta new cases trend data
 7. Daily mortuary rate with COVID-19 protocol or without protocol
 8. Positivitiy rate of daily COVID-19 diagnostic workup
 9. Data on law violation by companies
10. Data on Jakarta air quality and traffic information
11. Network graph of COVID-19
```

**Data on bed availability and referral system**

An innovation delivered by EIS is that this system help government in pandemic control by presenting actual data on bed occupancy and availability across hospitals, especially isolation wards. EIS integrate both government-owned and private-owned hospitals in Jakarta. This transpiration on data had been proven to prevent delay in referral and further improve survival of critical COVID-19 patients that needs ICU, PICU, or NICU (Fig 10).

Data presented on EIS including isolation ward, ICU with and without negative pressure, pediatric ICU, neonatal ICU, operating theater with negative pressure, and hemodialysis facilities for COVID-19. Data were directly inputted by hospitals into the system and being updated hourly. More detail data also presented (Fig 11) specifically for each hospital. This might be one of the first system to present almost actual data on bed occupancy in Indonesia.

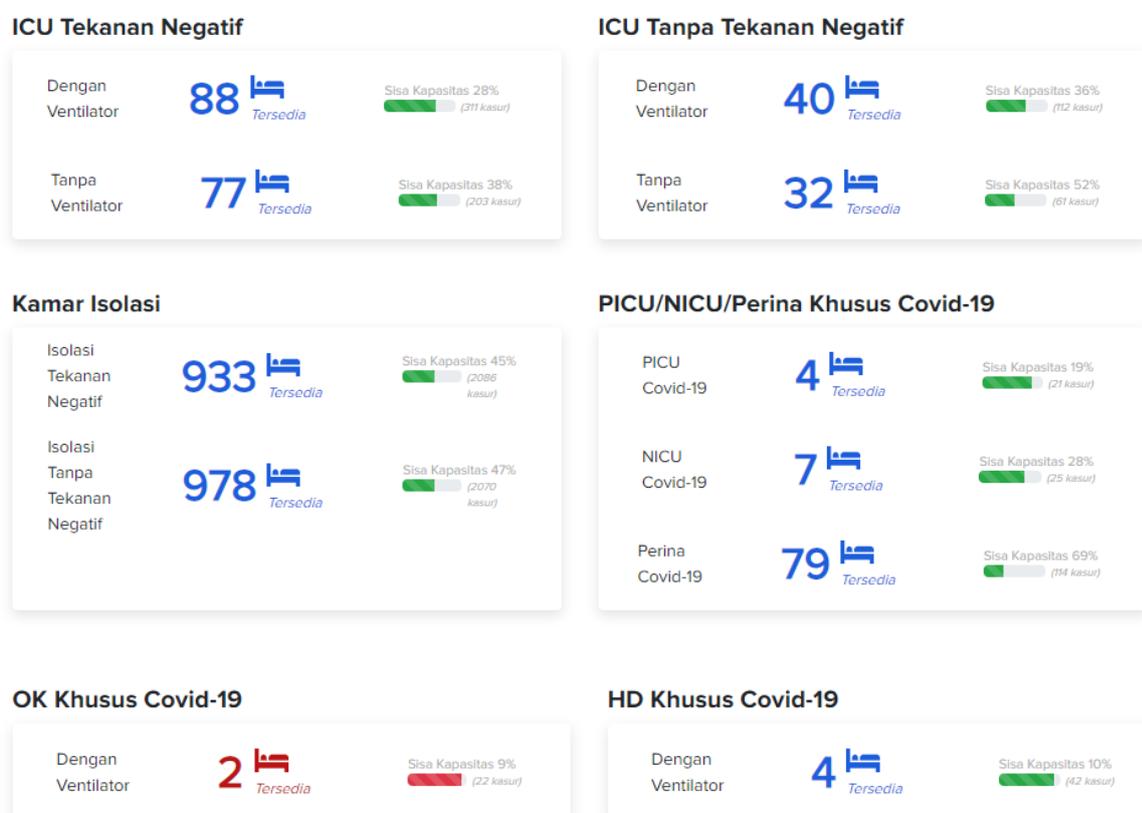

Figure 9. Bed Availability and Occupancy Data Presented in Executive Information System Jakarta. ICU tekanan negatif = negative pressure ICU; ICU tanpa tekanan negatif = without negative pressure ICU; Kamar isolasi = isolation room; OK Khusus COVID-19 = operating theater for COVID-19 cases; HD Khusus COVID-19 = hemodialysis facilities for COVID-19 cases; tersedia = availability in number; sisa kapasitas = capacity in percentage; dengan ventilator = with ventilator; tanpa ventilator = without ventilator

| No | Wilayah | Nama RS | Ketersediaan Tempat Tidur (Bed) | | | | | | PICU | NICU | Perina | OK |
|---|---|---|---|---|---|---|---|---|---|---|---|---|
| | | | ICU Tekanan Negatif | | ICU Tanpa Tekanan Negatif | | Isolasi | | | | | |
| | | | Dengan Ventilator | Tanpa Ventilator | Dengan Ventilator | Tanpa Ventilator | Tekanan Negatif | Tanpa Tekanan Negatif | | | | |
| 1 | Jakarta Pusat | RSUPN Dr. Cipto Mangunkusumo | 4 | 0 | 0 | 0 | 0 | 0 | 0 | 2 | 0 | 0 |
| 2 | Jakarta Pusat | RS Umum PAD Gatot Soebroto | 8 | 9 | 0 | 0 | 122 | 0 | 0 | 0 | 0 | 1 |
| 3 | Jakarta Pusat | RS Umum AL Dr Mintoharjo | 0 | 0 | 0 | 0 | 0 | 0 | 0 | 0 | 0 | 0 |
| 4 | Jakarta Pusat | RS Umum Daerah Tarakan | 7 | 22 | 0 | 0 | 28 | 0 | 0 | 0 | 3 | 0 |
| 5 | Jakarta Pusat | RS Umum Murni Teguh Sudirman Jakarta | 0 | 0 | 0 | 0 | 0 | 0 | 0 | 0 | 0 | 0 |
| 6 | Jakarta Pusat | RS PGI Cikini | 0 | 0 | 1 | 7 | 0 | 54 | 0 | 0 | 0 | 0 |

Figure 10. Bed Availability and Occupancy Status (per Hospital). Ketersediaan tempat tidur = bed availability; wilayah = region; nama RS = hospital name; ICU tekanan negatif = negative pressure ICU; ICU tanpa tekanan negatif = without negative pressure ICU; Kamar isolasi = isolation room; OK = operating theater; dengan ventilator = with ventilator; tanpa ventilator = without ventilator

## 3.  DISCUSSION

EIS brings Jakarta into a city with more developed data system on public health. This breakthrough should become an example for other cities of provinces in Indonesia and a further collaboration is always welcome. Some developing countries also have an information system to support policy making during COVID-19 pandemic. The key to fight pandemic is how to present data in transparency.[1,2]

Data reporting are a major issue in developing countries. India reported that only 10 out of 29 states that provide visual representation of COVID-19 cases trends, while most of states did not report data stratification in age, gender, district nor comorbidities.[3] Previous studies had reported that gender and age played an important role in COVID-19 data reporting due to its impact on severity and mortality.[4,5]

An objective indicator on how data reporting quality can be assessed using COVID-19 Data Reporting Score (CDRS) as developed by Stanford or other scoring systems, in which related to health development index.[3] This disparity would be common among developing countries, especially with geographical challenges like Jakarta, so that an intersectoral collaboration is necessary.

IT infrastructure and system also the major source of problem during system reporting. EIS might be easily implemented in Jakarta due to prior IT readiness, but a contrast finding could be met in other Indonesian cities. It needs a policy commitment to build an efficient IT system in a short time during this pandemic era.[6]

Some other concern that might impair data collection are about privacy. Some patients, both in developing and developed countries, are not giving permission of his civil ID number to be registered in national system as a COVID-19 cases. This phenomenon is strongly related to local social and spiritual belief, or a feeling of ashamed to be infected by SARS CoV-2. Thus, during EIS development, data privacy should be handled carefully to prevent further ethical issues.[7]

**Future direction**

The large amount of data, known as big data, should be optimized by academicians to publish scientific paper in both national and international level. Data utilization is not limited to health sectors, but social and economic institution can also bring this to a scientific evidence in order to help government for policy making.

EIS should also be integrated to central government system, such as New All Record Kemenkes, Peduli Lindungi application for contact tracing, self-isolation monitoring apps, and other new mobile-based apps. This collaboration can boost the quality of information extracted from the present data and open for any innovation in the future.

## REFERENCES AND CITATIONS